# Title: Full Automation for Rapid Modulator Characterization and Accurate Analysis Using SciPy

Authors:

T. L. Yap, A. Sasidhara, N. X. Ang, X. Guo, W. Wang, K. S. Ang, S. L. Tan

**Abstract**

Modulator testing involved complex biasing conditions, hardware connections and data analysis. Also, any optical signal distortion due to the grating coupler effect could potentially induce additional difficulty in setting the correct bias condition for an accurate measurement of the modulator performance. In this paper, we proposed to use SciPy, an open-source scientific computing library, for automation in the silicon modulator test with bias setting and data analysis.

**Introduction**

Wafer-level acceptance tests are essential for cost-effective mass production of silicon photonics (SiPh) products such as low power transceivers, LiDAR, sensors, etc. Wafer-level tests are performed on passive and active devices designed with grating couplers (GC). Good correlations between passive and active devices designed with GC and edge couplers (EC); often used in the final product; had been previously demonstrated [1]. However, due to different process steps between both sets of couplers, GC and EC might behave very differently when there is a big process drift or variation. This will increase the difficulty in extracting the actual performance of the modulator, independent of the light couplers. One of the common techniques used to eliminate the effect of light couplers is the use of normalization. However, this method is only effective on limited parameters in which the light coupler effect can be completely eliminated. One such example is propagation loss [1].

The modulator is one of the fundamental components in the optical communication network. Two of the most common silicon modulators are the Mach-Zehnder modulator (MZM) and the Micro-Ring modulator. The Micro-Ring modulator exhibits a natural resonant frequency based on the device dimension. Electro-Opto effect can be constructed by biasing PN junction on the micro-ring to change its refractive index. By changing the refractive index, its resonant frequency is shifted and thus resulting in a high modulation depth near the resonance [2]. On the other hand for MZM, by biasing PN junction on one or both arms of the Mach-Zehnder interferometer (MZI) will lead to changes in the refractive index on the arm(s) and the phase of light passing through it, effecting either constructive or deconstructive interference and consequently resulting in intensity modulation [2].

For both types of modulators to perform at the optimum modulation and bandwidth, the modulators need to be biased at the Quadrature point. The Quadrature point typically can be set either by tuning the heater bias (for test structure with a heater attached) or by tuning the laser wavelength (for unbalanced MZM or ring modulator). In this paper, the method used to set the Quadrature point using wavelength tuning during automatic testing will be discussed. SciPy is an ideal candidate for automation in silicon modulator biasing setting and data analysis, as the test executive used in this paper is written in Python programming language.

**Modulator Testing**

Figure 1 shows the overview of the algorithm flowchart CompoundTek test team developed for modulator testing. The wafer loading and alignment is handled by the prober. The DUT position on the wafer together with the location of the probe pads and grating couplers are setup through the prober file. The electrical probes and optical fiber can be moved to the designated pads and grating coupler position. After the electrical probes make contact with the pads, fine alignment of optical fibers to the grating couplers is required for optimum light coupling into the modulator. The fine optical alignment process is also known as peak search. After the peak search, the second part of the flow is handled by several test equipment, where the performance parameters of the modulators are measured

For accurate measurements, the polarization of the input light needs to be optimized before doing a wide wavelength sweep (~20nm) with 0V bias to the modulator. Free spectra range (FSR) can be calculated by identifying the notch (wavelength points with minimum light intensity). Upon completion, the modulator is biased up to the desired voltage condition. For each voltage condition, Quadrature point need to be set and the RF response bandwidth will be measured. ^ and † indicate the wavelength sliding method and curve fitting method to identify the Quadrature point. The methods will be described in details within the paper.

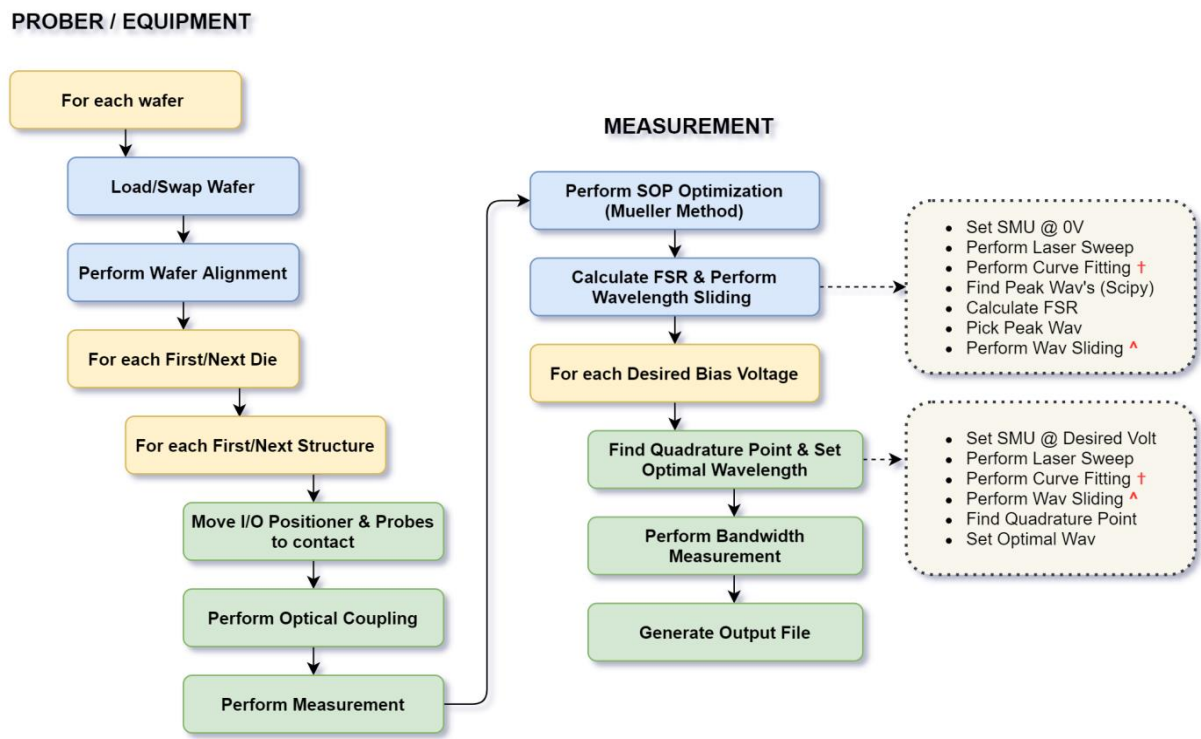

Figure 1 Algorithm flowchart developed by CompoundTek test team for modulator testing.

Figure 2 shows the schematics of the measurement setup for the modulator testing. The initial step is to set up an optical path by setting the laser at the wavelength of interest, tuned to an optimized polarization state. The light is then optically coupled in and out of the modulator, with the output connected through an optical switch to an optical power meter. Typically, for MZM testing, an optical amplifier is needed due to high insertion loss of the long MZI structure. After the optical path is established, the modulator electrical biasing condition is tuned to its linear operating region (Quadrature point). After which, the output from the modulator is then toggled, using the optical switch, from the optical power meter to the Lightwave Component Analyzer (LCA) for the 3-dB bandwidth measurement. The optical switch can be remotely controlled via GPIB (General Purpose Interface Bus to achieve a fully automatic, efficient and high precision photonics wafer test setup [1].

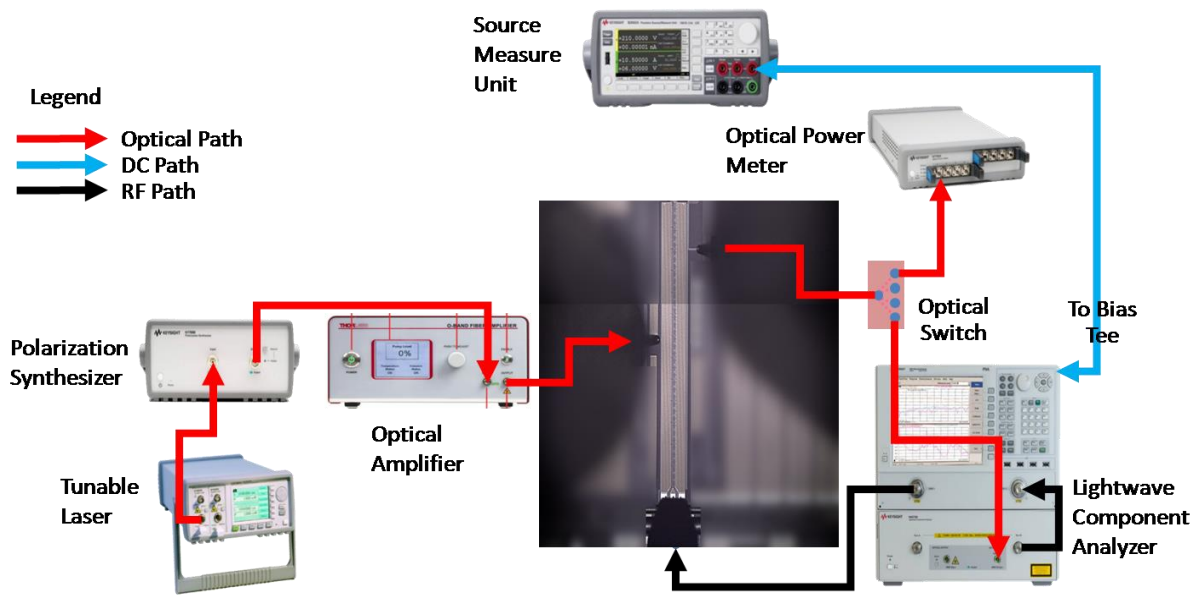

Figure 2: Schematics for silicon modulator testing

From Figure 3, the input signal from LCA can only be directly modulated without distortion, when the modulator is set to the Quadrature point (black curve). When the choice of wavelength is not optimized; either set at the left (blue curve) or right (red curve) of the optimized wavelength, the output modulated waveform curve will be distorted, affecting both the transmission characteristic curve and the measured bandwidth of the DUT modulator [3].

For balanced MZM with a heater attached to one of the output arms, the Quadrature point can also be set by tuning the heater bias. For unbalanced MZM or micro-ring modulator, the Quadrature point is usually set by tuning the wavelength of the optical signal. Discussion on methods to set the correct Quadrature point with wavelength tuning during automatic testing will be done in the later sections

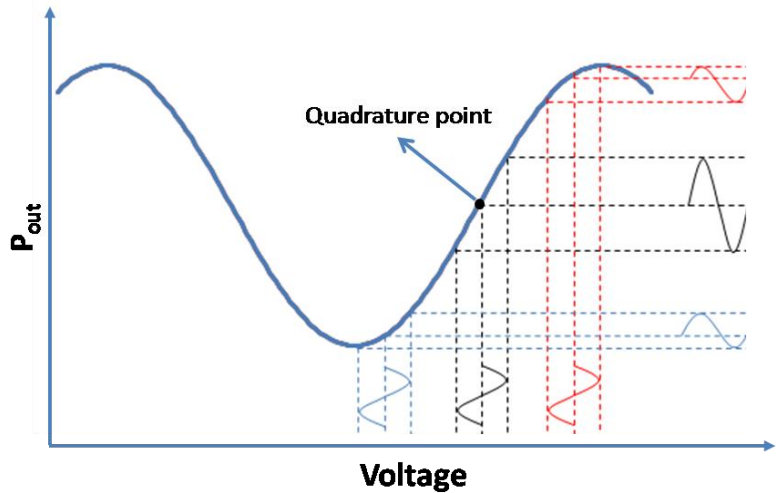

Figure 3: Typical optical spectrum of optical modulator with the Quadrature point. The effect of non-optimized choice of wavelength at either side of the Quadrature Point onthe distortion of output modulated waveform is illustrated. .

**Quadrature Point Setting with Wavelength Sliding**

The Quadrature point is usually set at 3dB lower from the maximum optical output. However, by performing just a simple search for $P_{max}$-3dB point, a wrong bias point might be found. This could be due to the optical spectrum being distorted by the Grating Coupler (GC) effect. Therefore, more logic and measures have to be implemented within the software coding to search for "true" Quadrature point. Wavelength sliding method is one of the methods which is highly effective to search for "true" Quadrature points and described below

From Figure 4, a wide wavelength spectrum is first sweep (20nm range) at 0V modulator bias to identify all the notches (red crosses on Figure 4a). The result is then used to calculate the free spectra range (FSR) and determine the notch closest to the targeted wavelength (1550nm for C band and 1310nm for O band, notch A, black dotted box on Figure 4a). Once the notch A is determined, the wavelength sweeping range is then slid to +/- 2nm from the notch A for subsequent sweep at desired modulator bias.  At each different modulator bias, a small shift in the notch is expected. Thus, an additional slide (+/- 2nm from the notch A) and sweep is needed at each different bias voltage (Figure 4b). With the narrower second range, we will be able to find the "true" Quadrature points.

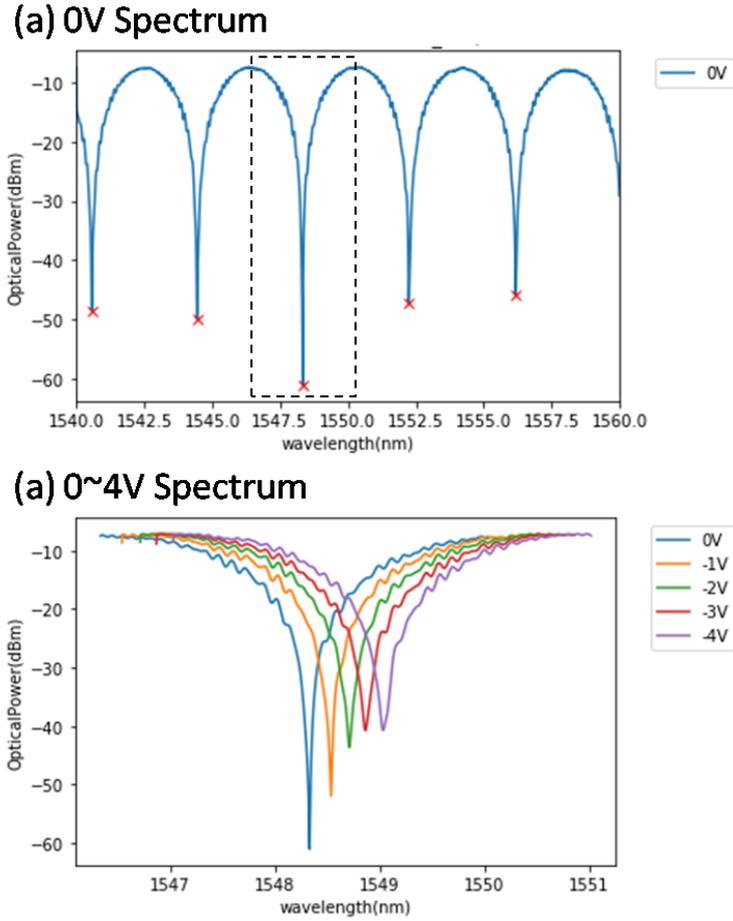

Figure 4 Illustration of wavelength sliding method. (a) wide spectrum @ 0V (b) narrow spectrum @ 0~4V for Quadrature point search.

**Quadrature Point Setting with Curve Fitting**

Wavelength sliding might not work as well if there are significant distortions to the narrower spectrums like high ripples. To resolve this, there is a need to go one step further to perform a curve fitting to the spectrum and use the fitted curve to search for Quadrature point. Fundamentally the relationship between the output intensity of the MZ modulator and the applied electric field voltage is as follows [5]:

$$P_{out} = P_{in} \left[ \frac{1 + \cos(\varphi_2 - \varphi_1)}{2} \right]$$

where $P_{out}$ and $P_{in}$ are the output and input optical power of the modulator, $\varphi_1$ and $\varphi_2$ are the phase shifts induced in its lower and upper branches of MZI, which are function of the optical wavelength. Therefore a nonlinear least squares can be used to fit the spectrum to a cosine function and the Quadrature point can be found from the fitted curve. The scipy.optimize.least_squares [4] is utilized for automation in the biasing setting. SciPy is an open-source scientific computing library in Python programming language and optimize.least_squares is a sub-package utilising de facto method 'lm' (Levenberg-Marquardt) to solve the nonlinear least-squares problem with bounds on the variables[6]. The residual function was set up as follow:

$$f_{res}(x, P_{out}; x_0, x_1, x_2, x_3) = x_0 \cdot \cos\left(\frac{2\pi}{x_1} \cdot x + x_2\right) + x_3 - P_{out}$$

where x is the wavelength and $P_{out}$ is the measured optical output. Scipy.optimize.least_squares finds the local minimum of $f_{res}$ by optimizing variables $x_0, x_1, x_2, x_3$. The optimized variables are also important parameter/Figure of Merit for the modulator. Physically $x_0$ is the extinction ratio (ER) of the modulator, $x_1$ is the free spectra range (FSR), $x_2$ is the phase change and $x_3$ is the $P_{max}$-3dB point.

On the other hand, resonant performance of micro-ring modulator can be accurately fitted with inversed Lorentzian function [7]. The residual function was set up as follow:

$$f_{res}(x, P_{out}; x_0, x_1, x_2, x_3) = \frac{-x_0}{\pi x_1 \left[1 + \left(\frac{x-x_2}{x_1}\right)^2\right]} + x_3 - P_{out}$$

where x is the wavelength and $P_{out}$ is the measured optical output

Physically $x_0$ is the ER, $x_2$ is the resonant frequency, $x_3$ is the $P_{max}$ point, and $x_1$ is the scale parameter which specifies the half-width at half-maximum (HWHM), alternatively $2x_1$ is full width at half maximum (FWHM).

**Conclusion**

Modulator is one of the key building blocks and components of Silicon Photonics product. However, testing of the modulator is sometimes complicated by the distortions introduced by other components such as the grating couplers, leading to incorrect results and wrong conclusions drawn. This paper describes how the wavelength sliding and curve fitting using SciPy can accurately and reliably find the correct condition the modulator must be set to get the correct bandwidth performance. This is important for the cost-effective mass production of SiPh products.